# Non-Flat Power Spectra in the CDM Model

Lutz V. Semig, Volker Müller

Astrophysikalisches Institut Potsdam, An der Sternwarte 16, D-14482 Potsdam, Germany

**Abstract.** Standard inflation with one scalar field produces primordial perturbations with a nearly flat ('Harrison-Zeldovich') power spectrum. Here we consider first, a double inflation spectrum, and second, a massive scalar field with an interaction potential which mimics an early quartic interaction, but fading away at a characteristic scale. We solve numerically the linear perturbation equations with initial conditions due to scalar field quantum fluctuations at the initial horizon crossing. The resulting power spectra are shown to be non-flat, exhibiting either a break or a valley. Using the transfer function of cold dark matter model we study the influence of the shape of primordial power spectra on observations of large scale structure in the universe. We compare the power spectra in redshift space with reconstructed power spectra from the IRAS catalogue. Further we discuss the variances of galaxy counts in cells, and the mass function of galaxy clusters. Comparison with standard CDM demonstrates the advantages and benefits of the more complicated initial spectra.

## 1. Introduction

The generation of the primordial perturbations responsible for structure formation is one of the most important questions in theoretical cosmology. Harrison (1970) and Zeldovich (1972) discussed some ad hoc ansatzes. A new basis was initiated with the development of the inflationary scenario. While inflation was initially designed to solve some longstanding cosmological paradoxa (Starobinsky 1980; Guth 1981; Linde 1983), it was realized soon that it provides an attractive solution to the problem of determining the initial perturbations for structure formation. Inflation makes definite predictions for the spectrum, in particular it is given by the form of the inflaton potential (Bardeen et al. 1983). Some authors claimed that for any desired spectrum, a corresponding potential can be constructed (Hodges & Blumenthal 1990). Essentially, this conclusion depends on the slow motion approximation, and it requires in general quite complicated potentials where the evolution may even contradict the slow motion approximation. Working in the range of power law potentials, non-trivial spectra are produced for example by a second inflaton field (Starobinsky 1985). Based on this idea, models for double inflation were constructed which are characterized by a *decreasing* of the power of scalar curvature fluctuations with rising wave numbers (Kofman & Linde 1987; Kofman & Pogosyan 1988; Starobinksy & Polarsky 1992). A related model takes into account $R^2$ modifications of the gravitational action (Gottlöber et al. 1991). A discontinuity in the gradient of the inflaton potential was suggested as cause for a special feature in the power spectrum at a typical scale (Starobinsky 1992, Ivanov et al. 1994). Here we propose a model with a valley at a given scale. To this aim we consider an inflaton potential with a quartic self-interaction at large values of the inflaton field, and an ordinary massive scalar field for small field values. This simple interpretation connects the break of the scale invariance of the primordial power spectrum with some phase transition in the scalar field theory. We take a simple parametrisation of a modified potential where the parameters are connected with the depth, the width and the place of the valley.

Further, we want to discuss consequences of modified primordial perturbation spectra for structure formation. To this aim, we need a transfer function describing the modification of the perturbation spectra through the epoch of equality and recombination. The standard CDM model with a primordial Harrison-Zeldovich spectrum was quite successful in describing the galaxy clustering on scales up to 20 $h^{-1}$Mpc. The transfer function is completely specified by a parameter $\Gamma = \Omega h$, where standard CDM means $\Omega = 1$ (the prediction of almost all inflationary models) and a dimensionless Hubble parameter $h = H_0/100 \text{km s}^{-1}\text{Mpc}^{-1} = 0.5$. Because of this success in the region of small scale galaxy distribution, we take the same transfer function. Standard CDM requires only one parameter, the normalization, which is now quite certainly determined by the COBE measure-ment (Smoot et al. 1992). A different measure of the amplitude of galaxy clustering is determined by biasing parameter relating the galaxy and dark matter distribution, observational, $b \sim 1.5\ldots 2$ (but this is not comparable with the COBE normalization, which requires $b \approx 1$).

It should be mentioned, that standard inflation with a polynomial potential already modifies the simple assumption of a scale invariant spectrum, due to the decreasing of the Hubble parameter during inflation, 'small' logarithmic corrections are imposed, which gradually reduce the power at small scales. The more dramatic effects discussed here are motivated by different reconstructions of the galaxy power spectrum, either using huge redshift surveys (Fisher et al. 1993; Park et al. 1994), the analysis of galaxy clustering on large scales (Loveday et al. 1992), or the study of galaxy clusters (cp. the discussion in Cen & Bahcall 1993).

The approximate scale invariance of the primordial perturbation spectrum is a quite stable property of the inflationary scenario, it is characteristic for all the new, the chaotic or the so-called natural inflation. But already an exponential inflationary potential (with very weak slope) leads to power law and not to quasi-exponential expansion during inflation, and in consequence to a tilt in the primordial perturbation spectrum. Observational consequences of this model are discussed by Cen & Ostriker (1992) and others. The 'tilt' requires only one additional parameter, but it has some shortcomings. In particular, it leads to the generation of primordial gravitational waves. Therefore, the COBE signal is not completely due to scalar potential fluctuations. Further, the power at small scales is strongly suppressed, leading to a very late formation of small structures in the universe, probably in contradiction with observations.

In general, any modification of the inflationary model changes essentially the scalar field dynamics and leads to strong deviations of the primordial power spectrum. We compare our spectra with the models of a broken scale invariance (BSI) discussed by Gottlöber et al. (1991). They are characterized by a step like power spectrum of the primordial gravitational potential. A similar spectrum is produced by two massive scalar fields (Polarski & Starobinsky 1993). These models are characterized by two additional parameters, the scale and the break height of the spectrum. Further, it was shown that any modifications of the power spectrum at the scales of galaxy clustering are strongly restricted by the relevant observations, for BSI, cp. Gottlöber et al. (1994), for two massive scalar fields, cp. Polarski (1994). The situation is similar to other modifications of the cosmological evolution, as for a mixture of cold and hot dark matter, or a tilted spectral index $n \approx 0.7$, or a model with $\Omega < 1.0$, without or with a cosmological constant $\Lambda$. It is typical for all these modifications that at small scales, the power of the perturbations is suppressed with respect to standard CDM. Therefore, all these models are characterized by a late formation epoch of small scale inhomogeneities.

Here we want to preserve the power of standard CDM on *both small and large* scales, suppressing the power only at intermediate scales. Therefore, we have to introduce a third parameter, the width of the suppression of the power. The proposed simple model has enough freedom for studying the observational consequences of different characteristics, in particular, the depth and the width of the suppression is connected with the steepness of the potential at the corresponding range. We will show, that for power spectra being in accordance with the reconstructed power spectrum say from the IRAS redshift survey, the slope of the spectrum at cluster scale is strongly restricted by the cluster abundance and the redshift evolution of clustering. The plan of the paper is as follows: Next we discuss the different inflationary models. In section 3, we summarize the method for calculating the primordial perturbation spectra. Section 4 is devoted to a discussion of the transformation of the primordial spectrum to the present time. All these calculations employ linear perturbation theory. In Section 5 we compare the linear perturbation spectrum with the power spectrum of the large scale galaxy distribution, essentially using the IRAS survey (Fisher et al. 1993). In Section 6 we study the abundance of deep potential wells responsible for galaxy clusters in the framework of the Press-Schechter theory. We conclude with an outlook of the parameter range allowed by observations for the discussed non-minimal inflationary models.

## 2. Inflationary models

The inflationary evolution is characterized by the dominance of the potential energy of a coherent scalar field $\varphi$ during the early cosmological evolution. The potential leads to an exponential growth in the scale factor, the first necessary step for explaining the origin of large scale structures. We consider a theory stemming from a general Lagrangian density $\mathcal{L}$ for gravity with a renormalization correction and one minimally coupled scalar field

$$\mathcal{L} = \frac{1}{16\pi G}\left(-R + \frac{R^2}{6M^2}\right) + \frac{1}{2}\varphi_{,\mu}\varphi^{,\mu} - V(\varphi), \qquad (1)$$

where $R$ is the scalar curvature, $\mu = 0,\ldots 3$, and we use units $c = \hbar = 1$, so that the Planck mass is the inverse of the square root of the Newtonian gravitational constant $G$, $m_{Pl} = G^{-1/2}$. The mass $M$ is the coupling constant of the $R^2$ term, being about four orders of magnitude smaller than $m_{Pl}$. We consider a spatially flat Friedmann-Robertson-Walker metric with scale factor $a(t)$. Then we have to solve the Klein-Gordon equation in this geometry,

$$\ddot{\varphi} + 3H\dot{\varphi} + V_{,\varphi} = 0, \qquad (2)$$

($H = \dot{a}/a$, the overdot means time derivative) and in the case of general relativity ($M^2 \to \infty$) the Friedmann equation with a scalar field energy density as source term,

$$H^2 = \frac{4\pi G}{3}\left(\dot{\varphi}^2 + 2V(\varphi)\right). \tag{3}$$

The equations for the complete Lagranian given above are discussed in Gottlöber et al. (1991). We take the more complicated terms resulting from the $R^2$ action and the resulting perturbation spectrum with t break at a characteristic scale as a reference.

We compare three different inflationary models and their influence on the power spectrum. First taking $M \to \infty$ in (1), we get typically a single inflationary stage. Here we employ a scalar field with mass $m$ and no selfinteraction,

$$V(\varphi) = \frac{m^2}{2}\varphi^2 \tag{4}$$

The background field $\varphi$ starts with an initial value $\varphi_0$, and it rolls down the potential to the minimum at $\varphi = 0$ where inflation stops. The total growth of the scale factor depends only on the initial value $\varphi_0$,

$$\ln(a_{fin}/a_0) = 2\pi G \varphi_0^2. \tag{5}$$

The solution of the horizon problem (and other cosmological paradoxa) requires a minimum value of 60 e-folds which sets a minimum to the initial value of the scalar field.

The considered double inflationary model (BSI model) takes $M \neq 0$ and also a massiv scalar field. The $R+R^2$-theory is conformally equivalent to general relativity with a second scalar field possessing a self-interacting potential (cp., e.g. Gottlöber et al. 1992)

$$V(\psi) = M^4(1 - \exp(-\psi/m_{PL}))^2. \tag{6}$$

Asymptotically ($\psi \to 0$), this corresponds to two scalar fields with different masses. The first inflationary phase is produced by the heavy field, it ends if the field reaches its potential minimum. Then starts the second inflation of the other field.

In the third model, we again use $M \to \infty$ but a special potential of changing self-interaction (dubbed CSI model, cp. Fig. 1)

$$V(\varphi) = \frac{m^2}{2}\varphi^2 + \frac{\lambda}{4}\varphi^4 \tanh(\alpha \frac{\varphi^2 - \beta^2 m_{Pl}^2}{m_{Pl}^2}) \tag{7}$$

where $\alpha$, $\beta$, and $\lambda$ are adjustable parameters. Initially we have a massive self-interacting scalar field evolving in time ($\varphi \gg \tilde{\varphi} \approx \beta m_{Pl}$). At the critical value ($\varphi \sim \tilde{\varphi}$) the self-interacting is switched off. The slope of this potential differs strongly from the quadratic potential near this critical value. With asymptotically vanishing interaction ($\varphi \ll \tilde{\varphi}$) the potential describes a pure massive scalar field. In any case, we determine the evolution of full scalar field dynamics, but despite the feature in the potential, the slow motion approximation holds true, since the potential is very flat. Checking the accuracy we get smaller than 2% deviation in the evolution of the scalar field between the exact theory and the slow motion approximation. In section 5 and 6 we deliver arguments for selecting special parameter values, $\alpha = 0.82$, $\beta = 2.45$, and $\lambda = 1.3 \times 10^{-8}$. Besides the dimensionless self-interaction constant, these are of order unity.

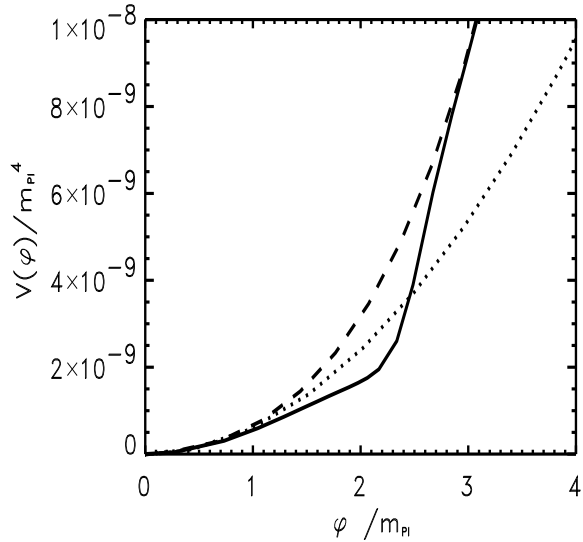

**Fig. 1.** Potential of the CSI model (solid line) compared with a massiv field with quartic self-interaction (dashed line) and a massive field without self-interaction (dotted line).

## 3. Perturbations

According to the inflationary scenario, quantum fluctuations in the scalar field and the metric are the basic mechanism for the perturbative deviations from homogeneity in the universe. Later the perturbations are stretched up to the observable large scale structure of the universe. To describe these fluctuations we have to decompose the scalar field into a homogeneous background part $\varphi$ and into small perturbations $\delta\varphi$. Here we give the equations in the case of one scalar field. Working in the longitudinal gauge where the metric includes two scalar perturbations $\Phi$ and $\Psi$, we have the line element

$$ds^2 = (1 + 2\Phi)dt^2 - a^2(t)(1 - 2\Psi)\delta_{ik}dx^i dx^k. \tag{8}$$

The equations of motion for the perturbations follow from the Klein-Gordon equation for the scalar field ($k$ is the wavevector)

$$\delta\ddot{\varphi} + 3H\delta\dot{\varphi} + [\frac{k^2}{a^2} + V_{,\varphi\varphi}]\delta\varphi = 4\dot{\varphi}\dot{\Phi} - 2V_{,\varphi}\Phi \qquad (9)$$

and the Einstein equations for the potential fluctuations (the spatial component $i \neq j$, $i, j = 1, 2, 3$)

$$\Phi = \Psi \qquad (10)$$

(and $0i$- component)

$$\dot{\Phi} + H\Phi = 4\pi G\dot{\varphi}\delta\varphi. \qquad (11)$$

To get the initial values for the perturbations we have to quantize the Klein-Gordon equation. During inflation the vacuum state is given by the deSitter space. With the asymptotic expansion $aH/k \to 0$, describing scales which are inside the Hubble radius, the quantization is as in the case for a flat space-time. The full quantization scheme is as discussed, for example, in the review of Mukhanov et al. (1992). Finally we have $\langle\delta\varphi^2\rangle^{1/2} = H/\sqrt{2}k^{3/2}|_{hor}$ and $\langle\delta\dot{\varphi}^2\rangle^{1/2} = \sqrt{2}H^2/k^{3/2}|_{hor}$, the suffix means the time of crossing the Hubble radius. As usual we neglect the index $k$ of the Fourier transformed quantities. The ordinary differential equations are solved numerically. In this way we get the primordial power spectrum of the potential perturbations $P_\Phi(k) = |\Phi|^2 k^3$.

A single massive field leads to a flat spectrum with a logarithmic correction (Fig. 2). This is approximately the Harrison-Zeldovich spectrum, and we take this model as the standard spectrum in the CDM calculations. The double inflationary model results in two different plateaus, each from the both inflationary stages. There results a step like spectrum, but with a wide transitionary stage and with superimposed small fluctuations at $k$-values after the break. The power spectrum for the potential (eq. 7) also decreases like the BSI model till $k \approx 1 \, h \, \text{Mpc}^{-1}$), but the dominance of the mass term at low field values leads to a secondary inflation which causes an increase of the spectrum. Some observational consequences of the three perturbation spectra will be discussed in the rest of the paper.

## 4. CDM and COBE normalization

In order to study consequences of the considered primordial power spectra for the large scale matter distribution, we have to describe the further evolution of the perturbations. After inflation, the modes reenter the Hubble radius, when starting from small scales, dissipative effects are to be taken into account. They depend on the kind of dark matter. Formally the different growth rates after reentering the horizon and the dissipative damping are described in a transfer function $T(k)$, relating the late time and the initial perturbation amplitude. As mentioned in the introduction, we assume the transfer function of CDM with shape parameter $\Gamma = \Omega h = 0.5$. A parametrization is given by (Bardeen et al. 1986), with $q = k/\Omega h^2 \text{Mpc}^{-1}$,

$$T(k) = \frac{\ln(1.0 + 2.34q)/2.34q}{[1 + 3.89q + (16.1 * q)^2 + (5.46q)^3 + (6.71q)^4]^{\frac{1}{4}}}. (12)$$

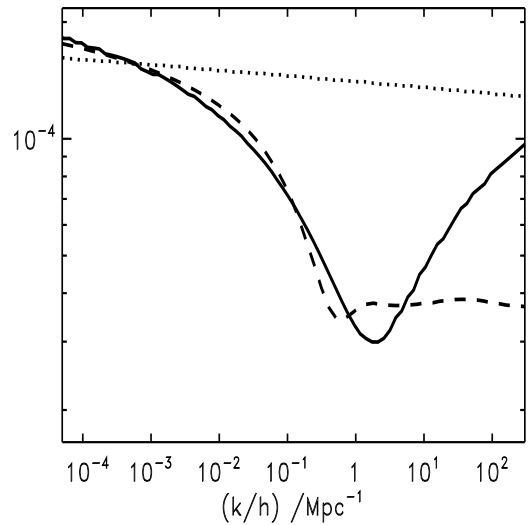

**Fig. 2.** The three different primordial power spectra from our models. For a massive field (dotted line) we get an approximate HZ spectrum. The BSI model (dashed line) produces a step in the spectrum, while the CSI model (solid line) results in a valley shaped spectrum.

For our inflationary models, we take $\Omega = 1$. From the Poisson equation there follows the connection of the potential with the density perturbations

$$P(k) = |\delta_k|^2 = \left(\frac{2}{3a^2H^2}\right)^2 T(k)^2 k^4 \Phi_k^2, \qquad (13)$$

where $\delta_k$ are the Fourier modes of the relative density perturbation $\delta\rho/\rho$. Finally we normalize this spectrum at the scale of the present horizon using the anisotropy $Q$ of the microwave background fluctuation. We employ a quadrupole normalization of the temperature fluctuation $\Delta T/T$

$$Q^2 = \frac{5T_0^2}{8\pi^2 R_H^4} \int_0^\infty dk \frac{P(k)}{k^2} j_2^2(2R_H k), \qquad (14)$$

where the Hubble radius $R_H = 3000 \, h^{-1}\text{Mpc}$ and $T_0 = 2.73$ K. The second year of $COBE$ data gives a value $Q = 19.9\mu$K (Bunn et al. 1994). Figure 3 shows the normalized dark matter power spectra for the SCDM (dotted line), the BSI (dashed line) and CSI (solid line) model. Already in this figure, one notes a decrease of the perturbation power at galaxy scales of the non-flat with respect to the standard CDM model.

In eq. (14), we neglected the contribution of gravitational waves, which in principle can become quite important for the non-flat perturbation spectra discussed in this paper. Taking into account only the Sachs-Wolfe effect, one gets for the ratio of the tensor to scalar contribution of the microwave background anisotropy, $Q_T/Q_S = 5m_P|dH/d\phi|/(8\pi)^{1/2}H$ (e.g. Lidsey & Coles (1992)). At large scales responsible for the quadrupole anisotropy, this ratio is about 0.58. Since the scalar and tensor contributions to the microwave background anisotropy add in quadrature, we get about a 37% reduction of the normalisation of the potential perturbations $P(k)$ in the CSI model, and in consequence a corresponding enlargement of the bias factor. The following discussion tests the form of the spectrum, therefore it is not influenced by this effect. The different scale dependences of scalar and tensor perturbations will be discussed in a separate paper.

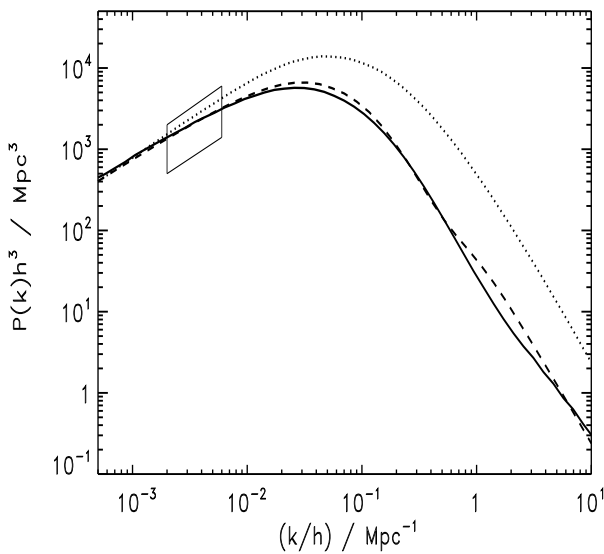

**Fig. 3.** COBE (box) normalized power spectra in the CDM model. The main advantage of the BSI (dashed line) and CSI (solid line) model compared with the Standard CDM (dotted line) model is less power on scales $k \geq 0.1\ h\ \text{Mpc}^{-1}$.

## 5. Comparison with Galaxy Observations

One has to distinguish between the dark matter and the galaxy distribution, since the density contrast of galaxies is higher. This circumstance is described by a linear bias factor

$$\left(\frac{\Delta\rho}{\rho}\right)_{gal} = b\left(\frac{\Delta\rho}{\rho}\right)_{DM}. \qquad (15)$$

We determine it by the dispersion in mass fluctuations for the different power spectra

$$\sigma^2(R) = \frac{1}{2\pi^2}\int_0^\infty dk k^2 P(k) W(kR) \qquad (16)$$

with a spherical 'top-hat' window function

$$W(kR) = 3\left(\frac{\sin(kR)}{(kR)^3} - \frac{\cos(kR)}{(kR)^2}\right). \qquad (17)$$

The CfA observations provide the result of variance unity of galaxies number counts in spheres of radius 8 $h^{-1}$Mpc, i.e. $\sigma_{gal}(8h^{-1}\text{Mpc}) = 1$ (Davis & Peebles, 1983). Then the bias factor is $b = 1/\sigma_{DM}(8\ h^{-1}\text{Mpc})$. For the standard CDM model, we get $b = 0.7$, and both for the BSI and the CSI model $b = 1.6$

In Figure 4 we show the comparison of variance measurements at scales larger than this normalization scale with our model curves. There we did not take into account a possible redshift correction which at theses scales amounts a constant scale-independent factor not affecting the form of the curves. The data points stem from the counts-in-cells analyses of Loveday et al. (1992) for the Stromlo APM survey, and from Efstathiou et al. (1990) using the IRAS catalogue. The standard CDM model gives too low values in the range of $20 - 60\ h^{-1}$Mpc, i.e. a too low level of inhomogeneities for the large scale clustering. The double inflationary model and the changing self-interaction spectrum show good agreement with the data.

Comparing theory with surveys in redshift space we have to include the peculiar velocities of galaxies in clusters by virialisation (Peacock 1991) and gravitationally induced streaming motions describing the infall of galaxies into overdensity regions (Kaiser 87). Fisher et al. (1993) give an approximation for both effects that we apply,

$$P_{red}(k) = \frac{f(\Omega,b) + (k/k_c)^2}{1 + (k/k_c)^2} P_{real}(k), \qquad (18)$$

with $k_c = 2\pi/20h^{-1}$ Mpc, and a bias dependent term

$$f(\Omega,b) = 1 + \frac{2}{3}\left(\frac{\Omega^{0.6}}{b}\right) + \frac{1}{5}\left(\frac{\Omega^{0.6}}{b}\right)^2. \qquad (19)$$

Putting them together and introducing an additional bias factor $b_{IRAS} = 0.7$ for comparison of optical with IRAS galaxies, we end up with a power spectrum in redshift space

$$P_{red}(k) = b_{Model}^2 b_{IRAS}^2 \frac{f(\Omega,b) + (k/k_c)^2}{1 + (k/k_c)^2} T^2(k) k^4 \Phi_k^2. \qquad (20)$$

The results are shown in Figure 5. We see that the BSI and CSI model both fit the $IRAS$ data (Fisher et al. 1993), in contrast to the CDM model with an approximate HZ spectrum.

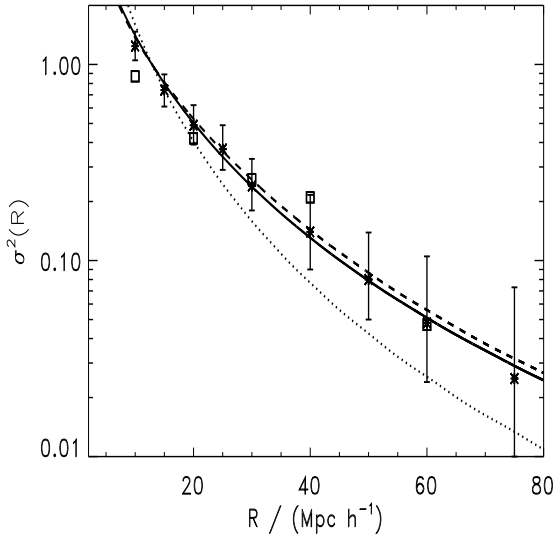

**Fig. 4.** Variance in the mass distribution from the counts-in-cells method of IRAS galaxies (Efstathiou et al. 90) (boxes) and the Stromlo APM survey (Loveday et al. 92) (stars). The different model curves correspond to the CSI model (solid line), the BSI model (dashed line), and the standard CDM model (dotted line).

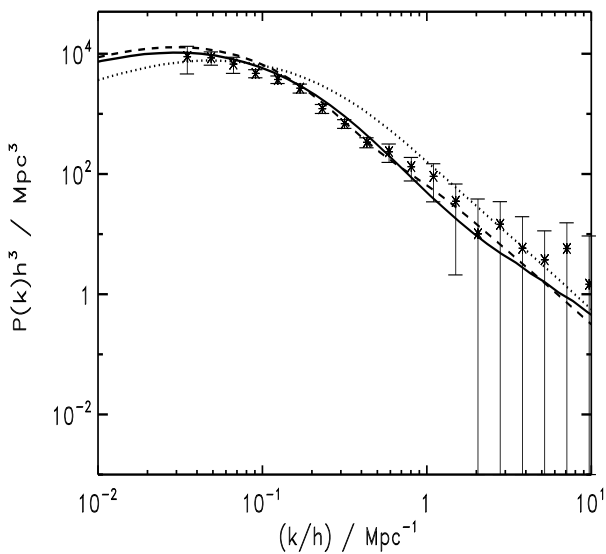

**Fig. 5.** The SCDM model (dotted line) has too much power compared to the IRAS galaxies (Fisher et al. 1993, the data are given in redshift space). BSI (dashed line) and CSI (solid line) models with the analytic redshift given in the text.

## 6. Mass function of clusters of galaxies

Clusters of galaxies are important cosmological probes, since they are the largest virialized objects in the universe. We calculate their abundance using the Press and Schechter (1974) theory and compare the result with optical and X-ray data. This phenomenological theory is based on one main assumption. If we smooth the density field $\delta$ by the spherical top-hat window function eq. (17) of scale $R$, then we expect a Gaussian probability of overdensity $\delta$,

$$P(\delta)d\delta = \frac{1}{\sqrt{2\pi\sigma(R)^2}} \exp(-\frac{\delta^2}{2\sigma(R)^2})d\delta. \quad (21)$$

For primordial quantum fluctuations and at scales where the linear theory of gravitational instability is applicable, this is a reasonable assumption. The essential step is the use of this probability also for condensed, i.e. nonlinear and presumable virialized objects. The dispersion $\sigma(R)$ is given in eq. (16). Using the assumption of spherical collapse which may be justified for high density contrast regions, one gets a critical overdensity $\delta = \delta_C = 1.69$ (e.g. Peebles 1980), where the formation of objects starts. The total number of bound objects follows from the integral $n(>M) = \int_{\delta_C}^{\infty} P(\delta, t)d\delta$. The mass function is given by the partial derivative $dn(>M)/dM$, we use the mean cosmic mass density to calculate the partial derivative in R instead. For a top-hat window function we have $M = 4\pi\bar{\rho}R^3/3$. To get the comoving number density we divide $n(M)$ by $M/\rho$ and multiply with 2 for normalization

$$n(M) = -\frac{1}{\sqrt{8\pi^3}} \frac{1}{M} \frac{\delta_C}{\sigma^2(R)} \frac{1}{R^2} \exp(-\frac{\delta_C^2}{2\sigma^2(R)}) \frac{d\sigma(R)}{dR}. \quad (22)$$

The fraction of bound objects with mass greater M is given by the number density $n(>M) = -\int_M^{\infty} n(\hat{M})d\hat{M}$.

In the wide mass range from $M \sim 10^{12}$ M$_\odot/h$ to $M \sim 10^{15}$ M$_\odot/h$, Bahcall & Cen (1993) estimated the cumulative mass function n(>M) of groups and clusters of galaxies (the smallest mass value is due to the estimate of the abundance of bright field galaxies). They used the richness class of Abell clusters including a standard luminosity of galaxies in clusters and an assumed mass-to-light ratio ($M/L = 300h$M$_\odot$/L$_\odot$) for estimating the cluster mass. The data are augmented by clusters with estimated known dispersion and some nearby cluster with detailed mass estimates. In addition Bahcall and Cen took the computer selected Edinburgh–Durham Cluster Catalogue (Lumsden at al. 1992), and further the X-ray temperature function of Henry & Arnaud (1991) for clusters from the Einstein satellite survey, assuming hydrostatic equilibrium for transforming the X-ray temperature to the mass. The data from the different sources are reproduced in Fig. 6. The number density distribution spans the wide range of $(0.01 \to 10^{-8})h^3$Mpc$^{-3}$, therefore it provides an important test of mass power spectra.

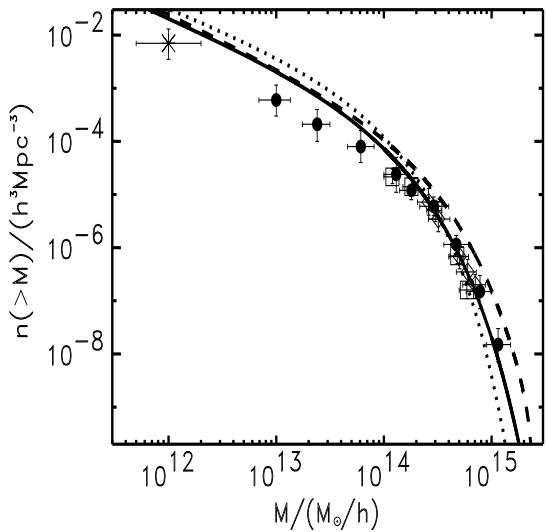

**Fig. 6.** Optical (dots and squares) and x-ray (diamond) data taken from Bahcall and Cen (1993). For comparison we give the results from the CSI potential (solid line), the BSI model (dashed line) and standard CDM (dotted line).

The mass functions for the three discussed models are given by the different curves in Fig. 6. The BSI and CSI results agree fairly well with the data. The fact that the CSI and BSI results are quite similar to each other is hardly astonishingly in fact of the similarity of both power spectra in the range responsibly for cluster formation. The standard CDM curve is steeper providing either a too high number of median and small mass halos or a too low probability of high mass clusters (this depends on the normalization of the curve, or, in other words, on the value of the critical overdensity in the Schechter formula determined by the bias factor). In fact, the cluster mass function provides a quite sensible test of the form of the spectrum. Indeed, the free parameters of the CSI model are quite strongly restricted by the cluster mass function, for such an estimate in the BSI model cp. Müller (1994). We impose a $\chi^2$ test for our valley like power spectrum. In principle, this may be problematic since the different points of the cumulative mass functions are not independent. But we believe, this restriction is not too severe in fact of the strongly decreasing mass function, i.e. effectively the low mass values are mainly determined by the low mass halo abundance. The $\chi^2$ test was used for different $k$ values corresponding to variations in the minimum of the primordial power spectrum. $k_M = 1.75\ h\mathrm{Mpc}^{-1}$ is the value used in comparing with the power spectra and the measured variances of counts in cells. There are two minima corresponding to $b = 1.75$, $k = k_M$ and $b = 1.97$, $k = 0.55 k_M$, they have a $\chi^2 = 27$, and 15, respectively (for 18 degrees of freedom). Even if the second minimum is somewhat deeper, it does not provide a fit to the $IRAS$ data in Fig.5. Therefore, we chose the first one as our preferred model. Our choice of $k_M$ follows from the sharp restriction of the effective power index at the scale responsible for cluster formation. The bias parameter must be fine tuned due to the exponential suppression of overcritical perturbations in the Press-Schechter theory.

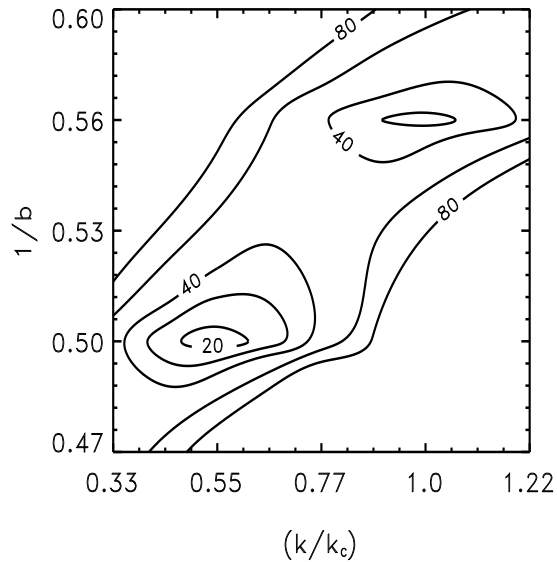

**Fig. 7.** Parameter space for the CSI model. The contour plot shows the allowed range judged by a $\chi^2$ test in dependence on the location of the minimum of the power spectrum $k_M$ and the bias. There are two minima. The parameters of the CSI potential are chosen to give $k_M = 1.75\ h\mathrm{Mpc}^{-1}$ and $b = 1.78$.

## 7. Conclusions

We discuss whether more complicated primordial perturbation spectra can ease some of the problems of the standard CDM model. To get a sufficiently wide range of possible 'break spectra', we introduced a changing self-interaction inflaton potential. The three parameters in the potential are strongly restricted by a range of large-scale structure observations in the universe. In each case, we started with COBE normalized spectra. Then already the requirement of a mildly nonlinear variance at a scale of $8h^{-1}\mathrm{Mpc}(\sigma \approx 0.5\ldots 1)$ restricts the *depth* of the suppression of the power. Reconstructed power spectra from galaxy surveys determine the *steepness* of the potential valley. Finally, the cluster mass function determines both the bias parameter and the *location* of the minimum of the gravitational potential. The redshift dependence of clustering, in particular, the abundance of the first condensed objects, determines the high $k$ asymptote of the spectrum. This question was not discussed in the present paper. It was our aim to demonstrate the strong restrictions of the primordial power spectra by observations of large-scale structures and to show how far any modifications of the standard CDM model are constrained. The introduced power spectrum with a 'valley' delivers an intersting example where these requirements become important, and it provides a quite good description of the different data sets.